\documentstyle[12pt]{article}
\def\be{\begin{equation}}
\def\ee{\end{equation}}
\begin{document}

\begin{titlepage}
{\hfill PUTP-94-10}
\begin{center}
\vspace {0.5 cm}
{\large\bf
The Quark Axial Vector Coupling and Heavy Meson Decays }
\vspace {0.5 cm}\\

{\bf Nan-Guang Chen}

Department of Physics, Peking University, Beijing 100871, P.R.China

\vspace {0.5 cm}

and

{\bf Kuang-Ta  Chao}

     CCAST (World Laboratory), Beijing 100080, P.R.China

	and

	Department of Physics, Peking University, Beijing 100871, P.R.China

\vspace {0.5 cm}
\end{center}
\vspace {0.5 cm}
\begin{abstract}

Form factors and decay widths for $D^{\ast}\rightarrow D \gamma$
and $D^{\ast}\rightarrow D \pi$ decays are estimated in
a relativistic constituent quark model. Relativistic corrections
due to light quarks are found to be substantial and to suppress
the vector and axial vector form factors. The $CLEO$ experimental
value of
$R^0_{\gamma}\equiv \Gamma (D^{\ast 0}\rightarrow D^{0} \gamma)/
\Gamma (D^{\ast 0}\rightarrow D^{0} \pi^{0})
=0.572\pm 0.057\pm 0.081$
is used to
determine the quark axial vector coupling $g_A$, which is found
to be $0.6-0.8$ for $m_u=(350-200) MeV$ correspondingly, as compared
with the chiral model result
$g_{A}= 0.8-0.9$. The heavy meson-pion strong
coupling $g$ is found to be $0.4-0.6$, much smaller than $g=1$
which is expected
in the large $N_{C}$ and nonrelativistic limit, but consistent with
some heavy hadron chiral theory and QCD sum rule results.
\end{abstract}

\end{titlepage}
\newpage

Because the chiral $SU(2)$ symmetry is spontaneously breaking, the
axial vector coupling of the constituent quark, $g_{A}$, may take any
values.
It has recently been argued based on a chiral model$^{[1]}$ that to
leading order in $1/N_{C}$
(where $N_{C}$ is the number of colors in quantum chromodynamics ),
the constituent quarks just behave like bare Dirac particles, and
both u and d quarks have vanishing anomalous magnetic moment and
have the axial vector coupling
\begin{equation}
g_{A}=1.
\end{equation}
With order $1/N_{C}$ corrections, the value of $g_{A}$ is slightly
reduced and is estimated to be$^{[2,3]}$
\begin{equation}
g_{A}=0.8-0.9.
\end{equation}
On the other hand, using the experimental value for the nucleon axial
vector coupling
$G_{A}=1.25$ with the static SU(6) wave function for the nucleon
leads to $g_{A}=1.25\times\frac{3}{5}=0.75$.
The discrepancy between this value and (2) might be removed by
including the relativistic corrections to the nucleon wave
functions$^{[4]}$.

The heavy meson decays, e.g., $D^{\ast}\rightarrow D $ decays may
also be a good testing ground for the value of quark axial vector
coupling
$g_{A}$. Although the total widths of $D^{\ast +}$ and $D^{\ast 0}$
are still unknown, their decay branching ratios have been measured
accurately by the $CLEO$ Collaboration$^{[5]}$ (see also ref.[6]).
These heavy
meson decays have been discussed in the literature$^{[7-13]}$. To
determine $g_{A}$, the most useful data might be
$B(D^{\ast 0}\rightarrow D^{0} \gamma)$ and
$B(D^{\ast 0}\rightarrow D^{0} \pi^{0})$, which are better measured
experimentally and are not sensitive to the value of the charm quark
mass in theoretical calculations. We suggest using the measured
ratio$^{[5]}$
\begin{equation}
R^0_{\gamma}\equiv \frac {\Gamma (D^{\ast 0}\rightarrow D^{0} \gamma)}%
{\Gamma (D^{\ast 0}\rightarrow D^{0} \pi^{0})}
=0.572\pm 0.057\pm 0.081,
\end{equation}
together with the calculated
$\Gamma (D^{\ast 0}\rightarrow D^{0} \gamma)$ and the form factors in
the hadronic decay to determine $g_{A}$.

Because the light constituent quarks ($u$ and $d$) inside the heavy
mesons are
relativistic, to discuss the $g_A$ issue in the heavy meson decays
the relativistic motion of light quarks has to be
taken into consideration.
In the following we will study the heavy meson decays and the $g_{A}$
problem in a relativistic constituent quark model, based on the
Bethe-Salpeter (BS) formalism.

For a $q_{1}\bar{q_2}$ bound system with quark momentum $q_1$, mass
$m_1$; antiquark momentum $q_2$, mass $m_2$; relative momentum q; and
total momentum $P$, meson mass $M$; and
\[p_{1}=\eta_{1}P+q, \  p_{2}=\eta_{2}P-q,\  \eta_{i}=\frac{m_i}
{m_{1}+m_{2}} \  (i=1,2), \]
the general form of the three dimensional BS wave functions for the
$0^{-}$ and $1^{-}$ mesons is given by (see e.g. ref.[14])
\begin{eqnarray}
\Phi^{P}_{P}(\vec{q})&=&\Lambda^{1}_{+}(\vec{p_1})\gamma^{0}
(1+\frac{\hat{P}}{M})\gamma_{5}\gamma^{0}\Lambda^{2}_{-}(\vec{p_2})
\phi (\vec{q})         \\
\Phi^{V}_{P}(\vec{q})&=&\Lambda^{1}_{+}(\vec{p_1})\gamma^{0}
(1+\frac{\hat{P}}{M})\hat{e}\gamma^{0}\Lambda^{2}_{-}(\vec{p_2})
f(\vec{q})
\end{eqnarray}
where $e_{\mu}$ is the polarization vector of the $1^{-}$ meson,
$e_{\mu}P^{\mu}=0; E_{i}=\sqrt{{\vec{p_{i}}}^{2}+m^{2}_{i}}$, and
$\Lambda^{i}_{\pm}=\frac{1}{2}(1\pm\frac{\gamma^{0}(\vec{\gamma}
\cdot\vec{p_{i}}+m_{i})}{E_{i}})$ are the positive (+) and negative
(-) energy projectors, and $\phi $, $f$ are scalar wave functions.
(4) and (5) respect the space reflection
symmetry,
and also respect the flavor-spin symmetry in the heavy quark limit.
This can be seen by taking $m_{1}\rightarrow\infty$, then $p^{\mu}_{1}
\rightarrow P^{\mu}$ and (4) and (5) become
\begin{eqnarray}
\Phi^{P}_{P}(\vec{q})&=&\frac{1}{v^{0}}
(1+\hat{v})\gamma_{5}\gamma^{0}\Lambda^{2}_{-}(\vec{p_2})
\phi (\vec{q}),         \\
\Phi^{V}_{P}(\vec{q})&=&\frac{1}{v^{0}}
(1+\hat{v})\hat{e}\gamma^{0}\Lambda^{2}_{-}(\vec{p_2})
f(\vec{q}),
\end{eqnarray}
where $v^{\mu}=\frac{P^{\mu}}{M}$, and $f=\phi$ which is due to the
vanishing of the color magnetic force in the heavy quark limit.
The normalization
condition for (4) and (5) is            \\
\begin{equation}
\frac{1}{2\pi}\int d^{3}q Tr\left[\Phi^{\dagger}_{P}(\vec{q})\Phi_{P}
(\vec{q})\right]=2P^0=2\sqrt{\vec{P}^{2}+M^{2}}.
\end{equation}
Suppose a flavor changing quark operator $\bar{q_f}\Gamma q_{i}$
induces the transition
\begin{equation}
\Phi_{P}(\vec{q})(p_{1},m{1};p_{2},m_{2};P,M)\rightarrow \Phi_{P'}
(\vec{q'})(p'_{1},m'_{1};p_{2},m_{2};P',M'),
\end{equation}
where the antiquark $(p_{2},m_{2})$ remains a spectator, then the
transition matrix element is given by
\begin{equation}
\langle P'\mid\bar{q_f}\Gamma q_{i}\mid P\rangle=
\frac{1}{2\pi}\int d^{3}p_{2}Tr\left[
\Phi^{\dagger}_{P'}(\vec{q'})\gamma^{0}\Gamma\Phi_{P}(\vec{q})\right].
\end{equation}
Likewise, the matrix element induced by the antiquark transition can
be written in a similar way.

In the radiative $M1$ decay such as $D^{\ast}\rightarrow D\gamma$ if
 we neglect the quark anomolous magnetic moment$^{[1,2]}$ we then
 only need to calculate the matrix element induced by  the vector
 operator $\bar{q}\gamma_{\mu}q$ of the quark and antiquark,
 and according to (10), (4), and (5), for the quark transition
we have
\begin{eqnarray}
j_{\mu}&=&\langle P'\mid\bar{q_1}\gamma_{\mu}q_{1}\mid P,e\rangle \\
\nonumber
&=&\frac{1}
{2\pi}\int d^{3}p_{2}Tr\left[\frac{\hat{p_2}-m_2}{2E_2}\gamma_{5}
(1+\hat{v'})\frac{\hat{p'_1}+m_1}{2E'_1}\gamma_{\mu}\frac{\hat{p_1}+m_1}
{2E_1}(1+\hat{v})\hat{e}\right] \\
\nonumber
& &\times f_{i}(\vec{p_{2}}-\frac{m_2}{m_{1}+m_{2}}\vec{P} )
        \phi_{f}(\vec{p_{2}}-\frac{m_2}{m_{1}+m_{2}}\vec{P'}   ),
\end{eqnarray}
where $\hat v=\frac{\hat{P}}{M_{i}},\hat{v'}=\frac{\hat{P'}}{M_{f}}$.
In general, the vector current matrix element (11) can be expressed
as
\begin{equation}
j_{\mu}= -i\sqrt{M_{i}M_{f}}\frac{M_{f}}{m_{1}}
\epsilon_{\mu\nu\alpha\beta}e^{\nu}v^{\alpha}v'^{\beta} \xi_{V1},
\end{equation}
where $\xi_{V1}$ is the vector form factor due to the quark $q_1$
transition. If the scalar wave functions $f_i$ (for the initial
state $1^-$ meson) and $\phi_f$ (for the final state $0^-$ meson)
in (11), which are to be determined by the interquark forces,
are known, then a direct calculation for (11) will give the
form factor $\xi_{V1}$. We will give this result later on by
solving the BS equation with a QCD-motivated interquark potential.
Before doing that calculation we may first consider a simplified
calculation as follows.
Because in the heavy quark limit the
color-magnetic force vanishes, the $0^{-}$ and $1^{-}$ mesons will
have the same spatial wave functions, we may assume $f_{i}$ and
$\phi_{f}$ to take the same form
\begin{equation}
f_{i}(\vec{p})=a_{i}f(\vec{p}),~~~~ \phi_{f}(\vec{p})=a_{f}f(\vec{p}),
\end{equation}
where $a_{i}$ and $a_{f}$ are normalization factors determined by (8)
in the initial and final meson frames respectively.
Although in (11) the light quark is
relativistic, in order to see the relativistic effects more
explicitly (but less rigorously) it might be
instructive to make a nonrelativistic expansion in terms of the
inverse of the quark masses. Then to the first order we find
\begin{equation}
\xi_{V1}=1-(\frac{2}
{3m^{2}_1}-\frac{1}{8m_{1}m_2})\langle\vec{q}^{2}\rangle
-\frac{1}{6}(\frac{m_2}{m_{1}+m_2})^{2}\mid\vec{K}\mid^{2}\langle\vec{r}
^{2}\rangle,
\end{equation}
where $\vec{K}=\vec{P'}-\vec{P}$ is the recoil momentum,
$\langle {\vec{q}}^{2} \rangle =\int d^{3}q{\mid f(\vec{q}) \mid}^2
{\vec{q}}^{2}$ is the mean value of the quark momentum squared,
and $\langle {\vec{r}}^{2}\rangle $ is the mean value of the radius
squared of the mesons.
{}From (14) we may find some qualitative feature of the relativistic
effects. We see that the M1
transition can be substantially suppressed if the quark is a light
quark and becomes
relativistic $\frac{\langle {\vec{q}}^{2} \rangle }{m^{2}_{1}} = O(1)$.
We see also that the effect of the spectator antiquark on the quark
transition
is not strong in general, because the coefficient of
the term involving $({m_1}{m_2})^{-1}$ is small.
The last term in (14) is due to the nonzero recoil momentum, and its
contribution is also small because of the smallness of the recoil momentum
in the $D^{\ast }$ radiative decays. If the quark is a heavy quark
then in the heavy quark limit $m_{1}\rightarrow \infty $
we will have $\xi_{V1}=1$ in (14) and
$\frac {M_{f}}{m_1}\rightarrow 1$ and therefore (12) will
return to the well known expression for the vector
current in the heavy quark limit.

In the rest frame of the $1^-$ meson, $\vec{v}=0, v^{0}=1, \vec{v'}=
\frac{-\vec{K}}{M_{f}}$, then (12) gives the familiar expression for the
M1 transition amplitude.
With a similar expression for the antiquark transition,
we can get the $1^{-}\rightarrow 0^{-}\gamma $ decay width for a
$q\bar{Q}$ system
\begin{equation}
\Gamma_{\gamma}=\frac{\alpha}{3}(\xi_{q}\frac{e_q}{m_q}+\xi_{Q}\frac
{e_Q}{m_Q})^{2}\mid \vec{K}\mid ^{3},
\end{equation}
where $\xi_{q}$ and $\xi_{Q}$ are the vector form factors $\xi _{V1}$
for $q=u,~d$ and $Q=c$ in the $D^{\ast }\rightarrow D\gamma$ decay.
We will calculate these form factors from (11) and will not use their
nonrelativistic expansion (14), because the light quark can be highly
relativistic.

The hadronic decay  $D^{\ast}\rightarrow D\pi$ can be described by the
quark-pion vertex in the chiral quark model$^{[15]}$
\begin{equation}
{\cal L}_{I}=-\frac{g_{A}}{2\sqrt{2}F_{\pi}}\bar{\psi}
\gamma^{\mu}\gamma_{5}\tau_{i}\psi\partial_{\mu}\pi^{i} ,
\end{equation}
where $g_{A}$ is the quark axial vector coupling and $F_{\pi}=132 MeV$
is the pion decay constant. In general, for a $q_{1}\bar {q_2}$ system
the transition induced by the quark axial
vector current can be written as
\begin{eqnarray}
j^{\mu}_{5}&=&\langle P'\mid\bar{q_1}\gamma^{\mu}\gamma_{5}q_{1}
\mid P,e\rangle \\ \nonumber
&=&\frac{1}
{2\pi}\int d^{3}p_{2}Tr\left[\frac{\hat{p_2}-m_2}{2E_2}\gamma_{5}
(1+\hat{v'})\frac{\hat{p'_1}+m_1}{2E'_1}\gamma^{\mu}\gamma_{5}
\frac{\hat{p_1}+m_1}{2E_1}(1+\hat{v})\hat{e}\right] \\ \nonumber
& &\times f_{i}(\vec{p_{2}}-\frac{m_2}{m_{1}+m_{2}}\vec{P} )
   \phi_{f}(\vec{p_{2}}-\frac{m_2}{m_{1}+m_{2}}\vec{P'})  \\ \nonumber
&=&\sqrt{M_{D^{\ast 0}}M_{D^{0}}}\left\{\xi_{A1}(1+v\cdot v')e^{\mu}-
\xi_{A2}(e\cdot v')v^{\mu}-\xi_{A3}(e\cdot v')v'^{\mu}\right\}.
\end{eqnarray}
As in the case of $\xi_{V1}$, we can calculate $\xi_{A1}, \xi_{A2}$,
and $\xi_{A3}$ with the scalar wave functions $f$ and $\phi$ obtained
by solving the BS equation, but it is useful to give
the nonrelativistic reduction form for, e.g., the $\xi_{A1}$
\begin{equation}
\xi_{A1}=1-\frac{\langle \vec{q}^{2} \rangle }{3 m^{2}_{1}}
-\frac{1}{6}(\frac{m_2}{m_{1}+m_2})^{2}\mid\vec{K}\mid^{2}\langle\vec{r}
^{2}\rangle.
\end{equation}
Again, $\xi_{A1}$ is suppressed by the relativistic motion of the
light quark, but the suppression is less severe than the M1
transition form factor $\xi_{V1}$ in (14). This can be seen by noting
that the coefficient of the
$\frac{\langle {\vec {q}}^{2}\rangle }{m^{2}_{1}}$ term, which is the
leading term for the suppression, is $-1/3$ in (18) whereas is $-2/3$
in (14).

For e.\ g.\ the $D^{\ast 0}\rightarrow D^{0}\pi^{0}$ decay,
$\mid\vec{P_{\pi}}
\mid =44MeV, \mid\vec{v'}\mid=0.024, \mid e\cdot v'\mid\ll 1,$ therefore
in (17) the
contribution of $\xi_{A2}$ and $\xi_{A3}$ terms can be neglected,
and we then get   \\
\begin{equation}
\Gamma(D^{\ast 0}\rightarrow D^{0}\pi^{0})=\frac{g^{2}_{A}\xi^{2}_{A1}}
{12\pi F^{2}_{\pi}}\mid\vec{P}_{\pi}\mid^{3}.
\end{equation}
For  $D^{\ast +}\rightarrow D^{+}\pi^{0}$ the width takes the same form
as (19), while for  $D^{\ast +}\rightarrow D^{0}\pi^{+}$ an additional
factor of 2 on the right hand side is needed due to the isospin difference.

To calculate the form factors in radiative decay (11) and (15) and in
hadronic decay (17) and (19), as a simple and naive choice,
we first use the Gaussian wave functions
\begin{equation}
f(\vec{q})=Ne^{-{\vec{q}}^{2}/a^{2}}
\end{equation}
for $f_i$ and $\phi_f$, where $N$ is the normalization factor,
$a^{2}=\frac{4}{3}\langle
{\vec{q}}^{2} \rangle$.
For the $D$ and $D^{\ast}$ mesons most estimates give
(see, e.g., ref.[16])
\begin{equation}
\langle {\vec{q}}^{2} \rangle=0.2-0.3~GeV^{2}
\end{equation}
Here we will simply take
$\langle {\vec{q}}^{2} \rangle=0.21~(0.30)~GeV^{2}$
for $m_u=m_d=200~(300)~MeV$, and $m_c=1500~MeV$,
and calculate the form factors using their relativistic
expressions (11) and (17) (not (14) and (18)). The calculated
form factors $\xi_{u}, \xi_{c},
\xi_{A1}$, and the decay widths are shown in Table 1. We see these
numerical results are qualitatively consistent with the nonrelativistic
reduction expressions (14) and (18) (e.g., $\xi_{u}$ is more suppressed
than $\xi_{A1}$). Here $g_{A}$ is
determined by using the  calculated
width for  $D^{\ast 0}\rightarrow D^{0}\gamma$ and the experimental
value for the ratio
$R^0_{\gamma}\equiv \frac {\Gamma (D^{\ast 0}\rightarrow D^{0} \gamma)}%
{\Gamma (D^{\ast 0}\rightarrow D^{0} \pi^{0})}
=0.572\pm 0.057\pm 0.081^{[5]}$.

To be more closely connected with QCD dynamics,
we have also calculated these form factors and decay widths based on the
BS equation with a QCD-motivated interquark potential$^{[14]}$.
In the instantaneous approximation the BS equation
\begin{equation}
( \not\!p_1-m_1) \chi _P(q) ( \not\!p%
_2+m_2) =\frac i{2\pi }\int d^4k~G(
\stackrel{\rightharpoonup }{P},
\stackrel{\rightharpoonup }{q}-\stackrel{\rightharpoonup }{k}) \chi
_P(k),
\end{equation}
where $G(\stackrel{\rightharpoonup
}{P},\stackrel{\rightharpoonup }{q}
-\stackrel{\rightharpoonup }{k})$ represents an
``instantaneous'' interquark potential in momentum space,
can be reduced to the following equation for the three dimensional
BS wave function
\begin{equation}
\Phi _{\stackrel{\rightharpoonup }{P}}( \stackrel{\rightharpoonup
}{q}%
) =\int dq^0\chi _P( q^0,\stackrel{\rightharpoonup }{q}),
\end{equation}
\begin{equation}
(P^0-E_1-E_2)\Phi _{\stackrel{\rightharpoonup }{P}}
(\stackrel{\rightharpoonup }{q}) = \Lambda _{+}^1\gamma ^0\int d^3k~G%
(\stackrel{\rightharpoonup }{P},\stackrel{\rightharpoonup }{q}-
\stackrel{\rightharpoonup }{k})
\Phi _{\stackrel{\rightharpoonup }{P}}( \stackrel{\rightharpoonup
}{k}) \gamma ^0\Lambda _{-}^2,
\end{equation}
where the contribution of negative energy projectors (i.e. the pair terms)
are neglected.

The interquark potential is described by a long-ranged
linear confining
potential ( Lorentz scalar $V_S$ ) plus a short-ranged one gluon
exchange potential
( Lorentz vector $V_V$ ), i.e.
\begin{eqnarray}
\label{a9}
&&{V(r)}={V_S(r)+\gamma_{\mu}\otimes\gamma^{\mu} V_V(r)},\nonumber \\
&&{V_S(r)}={\lambda r\frac {(1-e^{-\alpha r})}{\alpha
r}},\nonumber \\
&&{V_V(r)}=-{\frac 43}{\frac {\alpha_{s}(r)} r}e^{-\alpha r},
\end{eqnarray}
where the introduction of the factor $e^{-\alpha r}$ is to regulate
the infrared (IR) divergence and also to incorporate
the color screening
effects of the dynamical light quark pairs on the Q\=Q potential.
It is clear that when $\alpha r\ll 1$ the potentials given here
become identical with the standard linear plus Coulomb
potential. In momentum space the potentials are
\begin{eqnarray}
&&G( \stackrel{\rightharpoonup }{p})=G_S( \stackrel{%
\rightharpoonup }{p}) +\gamma_{\mu}\otimes \gamma^{\mu}
G_V( \stackrel{\rightharpoonup
}{p}),\nonumber \\
&&G_S( \stackrel{\rightharpoonup }{p})=-\frac \lambda \alpha
\delta ^3( \stackrel{\rightharpoonup }{p})+\frac \lambda {\pi
^2}\frac 1{( \stackrel{\rightharpoonup }{p}^2+\alpha ^2)
^2},\nonumber \\
&&G_V( \stackrel{\rightharpoonup }{p})=-\frac 2{3\pi^2}
\frac {\alpha_{s}(\stackrel{\rightharpoonup
}{p})}{\stackrel{\rightharpoonup }{p}^2+\alpha ^2},
\end{eqnarray}
where $\alpha_{s}(\stackrel{\rightharpoonup }{p})$ is the well known
running
coupling constant and is assumed to become a constant of $O(1)$ as
${\stackrel{\rightharpoonup }{p}}^2\rightarrow 0$
\begin{equation}
\alpha _s( \stackrel{\rightharpoonup }{p}) =\frac{12\pi }{27}%
\frac 1{\ln ( a+\frac{\stackrel{\rightharpoonup }{p}^2}{\Lambda
_{QCD}^2%
}) }.
\end{equation}
The constants $\lambda$, $\alpha$, $a$ and $\Lambda_{QCD}$ are
the parameters that characterize the potential.
In the computation we will use
$\lambda=0.18 GeV^{2}, ~~\alpha=0.06
GeV,~~a=e=2.7183,~~\Lambda_{QCD}=0.15 GeV$.

Substituting (4), (5), and (26), (27) into the reduced BS equation
(24), with quark masses $m_u=m_d=200-350~MeV,~m_c=1500~MeV$, we can
solve for the scalar wave functions $\phi $ and $f$
of the $0^-$ and $1^-$ mesons respectively,
which are different due to the color magnetic force induced by one
gluon exchange potential. We then substitute the obtained $\phi$
and $f$ into (11) and (17) to calculate the form factors.
The results are also shown in Table 1.

The form factors are more suppressed for the wave functions
by solving BS equation than for the Gaussian wave functions.
This is mainly because, due to the relativistic
correction of one gluon exchange potential (like the Breit-Fermi
Hamiltonian) in the BS equation the spatial wave function of the
$0^{-}$ meson has a larger $\langle {\vec{q}}^{2} \rangle $ (caused
by an attractive spin-spin force between the quark and antiquark),
and the $1^{-}$ meson wave function
becomes different from the $0^{-}$ meson wave function, and
therefore the overlap integral of wave functions between $D^{*}$
and $D$ is reduced. From Table 1 we see that although the
decay widths are predicted to be somewhat different in the two
models, the obtained values for $g_{A}$ are close
to each other. E.g., for $m_u=300 MeV$, $g_A=0.65-0.67$;
for $m_{u}=200 MeV$, $g_{A}=0.81-0.83$. Nevertheless, we prefer
the results obtained from BS equation with QCD-motivated potentials.

In the heavy quark limit $(m_{Q}\rightarrow\infty)$
for the hadronic decay $D^{\ast}\rightarrow D\pi$
the heavy meson-pion strong coupling $G_{D^{\ast}D\pi}$, defined by
\begin{equation}
\langle D^{0}(p_{f})\pi^{+}(p_{\pi})\mid D^{\ast +}(p_{i},\epsilon)
\rangle=G_{D^{\ast}D\pi}p^{\mu}_{\pi}\epsilon_{\mu},
\end{equation}
may be written in a more convenient form as
\begin{equation}
G_{D^{\ast}D\pi}=\frac {2M_{D^{\ast}}}{F_{\pi}}g.
\end{equation}
Comparing (28),(29), with (19) it is easy to find
\begin{equation}
g=g_{A}\xi_{A1}.
\end{equation}
In the large $N_{C}$ limit ( $g_{A}$=1 ) and the nonrelativistic
limit ($\xi_{A1}$=1 ) we would expect
\begin{equation}
g=1.
\end{equation}
However, based on the relativistic description for heavy meson
decays, with a typical value
of the constituent quark mass $m_{u}=350(300)MeV$ our BS model gives
\begin{equation}
\xi_{A1}=0.67(0.64),~~~~ g_{A}=0.60(0.65),~~~~ g=0.40(0.42).
\end{equation}
We see that our estimated values of $g_{A}$ are significantly smaller
than (2): $g_{A}=0.8-0.9$, which is expected in the chiral lagrangian
approach. This is similar to the result of ref.[12].
Moreover, our value for the meson strong coupling
$g\approx 0.4$ is also much smaller than (31): $g=1$.

However, our result for the quark axial vector coupling $g_{A}$ is
sensitive to the value of the constituent quark mass. We see from
Table 1 that if $m_u=200MeV$ then the BS model calculation would give
$\xi_{A1}=0.58, ~g_{A}=0.81.$ In this case,
although the light quark inside
the heavy meson becomes even more relativistic, the $M1$ transition
$D^{\ast 0}\rightarrow D^{0}\gamma$ width gets enhanced due to a larger
Dirac moment of the even lighter quark. Consequently, with the $CLEO$
ratio (3)
the strong decay width of $D^{\ast 0}$ meson and the value
of $g_A$ will be increased. This possibility
is of course not excluded and worth further investigating. In this
connection it might be interesting to notice that $m_u\approx 200~MeV$
is also suggested in some relativistic quark models (see, e.g., ref.[4]).

As for the meson strong coupling $g$, with $m_u=200MeV$
we get $g=0.47$, which is still much smaller than $g=1$.
In fact, our BS model calculations show that with
a wide range of the constituent quark mass, say, $m_u=200-350 MeV$, the
obtained meson strong coupling is $g=0.47-0.40$. The Gaussian wave
functions give $g=0.53-0.45$ for $m_u=200-300MeV$ (see Table 1).
All these results suggest
\begin{equation}
g=0.4-0.6,
\end{equation}
which is much smaller
than $g=1$ expected in the large $N_{C}$ and nonrelativistic limit,
and than $g=0.8-1$ suggested e.g. in refs.[8,19], but is consistent
with the result $g\approx 0.6$ in a heavy hadron chiral model$^{[9]}$.
It is interesting
to note that some recent QCD sum rule analyses favor an even smaller
value $g=0.2-0.4^{[17]}$. The study of semileptonic decay
$D\rightarrow\pi l\nu_{l}$ using a chiral effective theory in the
heavy quark limit also favors $g=0.4$$^{[18]}$.
Moreover, $g=\frac 13$ is suggested
by some calculation for the relativistic effects$^{[13]}$.
With (29) and (33) we obtain the following estimate for the effective
$D^{*}D\pi$ coupling and $B^{*}B\pi$ coupling
\begin{equation}
G_{D^{*}D\pi}=12-18,~~~~G_{B^{*}B\pi}=32-48.
\end{equation}

Finally, one might ask whether the nonrelativistic treatment for the
light quarks is still
tenable for heavy meson decays. In the nonrelativistic limit we
would have $\xi_u=\xi_c=\xi_{A1}=1$, then from (15) and the $CLEO$
ratio (3) with typical quark masses $m_c=1500~MeV,~m_u=m_d=330~MeV$
we would get $\Gamma_{tot}(D^{*0})=107~KeV,~\Gamma_{tot}(D^{*+})
=151KeV$, which already exceeds the observed upper bound
$\Gamma_{tot}(D^{*+})\langle 131~KeV^{[6]}$.
This seems to rule out the possibility that the light constituent
quark inside the heavy meson can be treated as nonrelativistic,
indicating that a relativistic description for the light quarks in
heavy meson decays is necessary.
It is amusing that by above nonrelativistic treatment we would get
\begin{equation}
g_{A}=0.72,
\end{equation}
which is almost in coincidence with the nonrelativistic value
$g_{A}=0.75$ obtained from the nucleon $\beta $-decay.

To sum up, we have estimated the form factors and decay rates for
the $D^{\ast}\rightarrow D\gamma$ and $D^{\ast}\rightarrow D\pi$
decays, based on a relativistic constituent quark model with QCD
inspired interquark potentials. Relativistic effects are found to
be substantial on the form factors. Using the $CLEO$ experimental value
$R^0_{\gamma}\equiv \frac {\Gamma (D^{\ast 0}\rightarrow D^{0} \gamma)}%
{\Gamma (D^{\ast 0}\rightarrow D^{0} \pi^{0})}
=0.572\pm 0.057\pm 0.081$ as input we find
that the quark axial vector coupling
might be consistent with $g_{A}=0.8-0.9$, which is expected in the
chiral lagrangian approach. However, this can only be achieved
with a smaller light quark mass, say, $m_{u}=m_{d}=(200-220) MeV$,
which give larger $M1$ transition widths accordingly.
With the typical value $m_{u}=m_{d}=300-350 MeV$, our obtained value for
$g_{A}$ is smaller, say $0.67-0.60$. It is therefore interesting to note
that the $g_{A}$ issue might be possibly related to the value of the
constituent quark mass of light quarks in the heavy meson decays.
As for the heavy meson-pion strong coupling $g$, with a
wide range of the constituent quark mass $m_u=m_d=(200-350)MeV$,
we obtain $g=0.6-0.4$, which is
much smaller than $g=1$ expected in the large $N_{C}$ and
nonrelativistic limit.



$
\begin{array}{|c|c|c|c|c|c|}
\hline
 & \multicolumn{2}{|c|}{Gaussian} & \multicolumn{3}{|c|}{BS} \\
 & m_u=200  & m_u=300  & m_u=200  & m_u=300  & m_u=350 \\
 &  (MeV)   &  (MeV)   &  (MeV)   &  (MeV)   &  (MeV)  \\
 \hline
 \xi _u & 0.42 & 0.51 & 0.36 & 0.46 & 0.51 \\
 \xi _c & 0.97 & 0.95 & 0.90 & 0.87 & 0.86 \\
 \xi _{A1} & 0.64 & 0.70 & 0.58 & 0.64 & 0.67 \\
 g_A &   0.83 & 0.65 & 0.81 & 0.65 & 0.60 \\
 g   &   0.53 & 0.45 & 0.47 & 0.42 & 0.40 \\
 \Gamma (D^{*0}\rightarrow D^0\gamma ) & 21(KeV) & 15(KeV) & 16(KeV) &
13(KeV) & 12(KeV) \\
 \Gamma (D^{*0}\rightarrow D^0\pi ^0) & 37 & 26 & 29 & 22 & 21 \\
 \Gamma_{tot} (D^{*0}) & 58 & 41 & 45 & 35 & 33 \\
 \Gamma (D^{*+}\rightarrow D^{+}\gamma ) & 0.42 & 0.12 & 0.24 & 0.10 &
0.07 \\
 \Gamma (D^{*+}\rightarrow D^{+}\pi ^0) &   26 & 18 & 20 & 15 & 14 \\
 \Gamma (D^{*+}\rightarrow D^0\pi ^{+}) &   55 & 40 & 43 & 33 & 31 \\
 \Gamma_{tot} (D^{*+}) & 81 & 58 & 63 & 48 & 46 \\
 \hline
\end{array}
$

{\parindent=0pt Table 1: Form factors
and decay widths for $D^{\ast}\rightarrow
D\gamma  $ and $  D^{\ast}\rightarrow D\pi$ decays. Predicted
values are given with (1) the Gaussian  wave functions
and (2) the wave functions by solving BS equation.
Here  $m_{c}=1500 MeV,
 m_{u}=m_{d}=200,~ 300,~350~ MeV~$ are assumed.  The CLEO experimental
value of
$R^0_{\gamma}\equiv \frac {\Gamma (D^{\ast 0}\rightarrow D^{0} \gamma)}%
{\Gamma (D^{\ast 0}\rightarrow D^{0} \pi^{0})}
=0.572\pm 0.057\pm 0.081$ is used as input.}

\bigskip

This work was supported in part by the National Natural Science Foundation
of China, and the State Education Commission of China. After this work
was submitted for publication we received the Bari preprints (refs.[13,17]),
and one of us (K.T.C) would like to thank Dr.De Fazio and Prof.Nardulli
for the useful communication.\\

\end{document}